\documentclass[useAMS,usenatbib]{mn2e}
\usepackage{epsfig}
\usepackage{natbib}
\usepackage{color}
\usepackage{graphicx}
\usepackage{amsmath}
\usepackage{amssymb}
\usepackage{natbib}
\usepackage{url}

\newcommand  \acc     {\ifmmode {\rm km\,s}^{-2} \else
km\,s$^{-2}$\fi}
\newcommand  \kms      {km\,s$^{-1}$}

\newcommand  \ergs     {\ifmmode {\rm ergs\,s}^{-1} \else ergs
s$^{-1}$\fi}
\newcommand  \ergcms   {\ifmmode {\rm erg~cm}^{-2}\,{\rm s}^{-1}
                        \else erg~cm$^{-2}$\,s$^{-1}$\fi}
\newcommand  \ergcmsA  {\ifmmode{\rm erg\,cm}^{-2}\,{\rm
s}^{-1}\,{\rm\AA}^{-1}
                        \else
ergs\,cm$^{-2}$\,s$^{-1}$\,\AA$^{-1}$\fi}
\newcommand  \ergcmsHz {\ifmmode{\rm ergs\,cm}^{-2}\,{\rm
s}^{-1}\,{\rm Hz}^{-1}
                        \else
ergs\,cm$^{-2}$\,s$^{-1}$\,Hz$^{-1}$\fi}
\newcommand  \phcms    {\ifmmode {\rm ph\,cm}^{-2}\,{\rm s}^{-1}
                        \else ph\,cm$^{-2}$\,s$^{-1}$\fi}
\newcommand  \phcmsA   {\ifmmode {\rm ph\,cm}^{-2}\,{\rm
s}^{-1}\,{\rm\AA}^{-1}
                        \else
ph\,cm$^{-2}$\,s$^{-1}$\,\AA$^{-1}$\fi}

\newcommand\aj{{AJ}}%
\newcommand\araa{{ARA\&A}}%
\newcommand\apj{{ApJ}}%
\newcommand\apjl{{ApJ}}%
\newcommand\apjs{{ApJS}}%
\newcommand\aap{{A\&A}}%
\newcommand\nar{{New Astronomy Reviews}}%
\newcommand\mnras{{MNRAS}}%
\newcommand\ssr{{Space~Sci.~Rev.}}%
\newcommand\nat{{Nature}}%
\title[White dwarf variability from Kepler]
{{\it Kepler} and the seven dwarfs: detection of
low-level day-timescale periodic
photometric variations in white dwarfs}
\author[D. Maoz, T. Mazeh, A. McQuillan]
{Dan Maoz, Tsevi Mazeh, Amy McQuillan\\
School of Physics and Astronomy, Tel-Aviv University,
Tel-Aviv 69978,
Israel} \date{\today}

\begin{document}

\maketitle

\label{firstpage}

\begin{abstract}
We make use of the high photometric precision of {\it Kepler} to
search for periodic
 modulations among 14 normal (DA- and DB-type, likely
non-magnetic) hot
 white dwarfs (WDs). 
In five, and possibly up to
seven of the WDs, we detect periodic, $\sim 2$~hr to 10~d,
variations, with
semi-amplitudes of 60--2000 ppm, lower than ever seen in WDs.
We consider various explanations:
 WD rotation combined with
magnetic cool spots; rotation combined with magnetic dichroism;
rotation combined with hot spots from an interstellar-medium
accretion flow;
transits by size $\sim 50$--$200$~km objects;
relativistic beaming due to reflex motion caused by a cool
companion WD; or reflection/re-radiation
of the primary WD light by a brown-dwarf or giant-planet companion,
undergoing
illumination phases as it orbits the WD. Each
mechanism could be behind some of the variable WDs, but could not be 
responsible for all five to seven variable cases. 
Alternatively, the periodicity
may arise from 
UV metal-line opacity,
associated with accretion of rocky material, a phenomenon seen in  
$\sim 50$\% of hot WDs. Non-uniform UV opacity, combined with 
WD rotation and 
fluorescent optical re-emission of the absorbed UV energy, could
 perhaps explain our findings. 
Even if
reflection by a planet is the cause in only a few
of the
 seven cases, it would imply that hot
Jupiters are very
common around WDs. 
If some of the rotation-related
mechanisms
are at work, then normal WDs rotate as slowly
as do peculiar WDs, the only kind for which precise
rotation measurements
have been
possible to date.
 \end{abstract}




\section{Introduction}
The final stage of stellar evolution, in which an
intermediate-mass
star loses a significant fraction of its mass and becomes a white
dwarf (WD),
is still incompletely understood theoretically, and poorly
constrained
observationally. Photometric variability can inform the problem
by revealing
WD rotation and WD companions --- planetary, substellar, or
stellar remnants.

For normal stars, the high photometric precision of
the {\it Kepler} mission has revolutionised our knowledge
of both companions and rotation. By detecting planetary transits,
{\it Kepler}
has discovered over 3500 planet candidates \citep{batalha13}, and
more than 2000 eclipsing
binaries \citep{slawson11}.
The BEER project \citep{faigler11}
is finding tens of companions to {\it Kepler} stars
through photometric modulation due to combinations of
relativistic beaming, tidally induced ellipticity, and reflection
of the
companion's light \citep{faigler12}.
Rotation periods have been measured
for 34,000 main-sequence stars, based on star-spot-induced
photometric
modulation \citep{mcquillan14}.

In this paper, we attempt to apply {\it Kepler}'s unique
capabilities to probe
for rotation and companions  in WDs.
We briefly review the observational state of the art
regarding each of these phenomena.

\subsection{WD rotation}
\label{sec:wdrotation}
 The rotation speed of a WD is one of the few remnant clues to
the physics of WD formation. If angular momentum were conserved
during the
formation of a degenerate stellar core, one would expect a factor
of
$\sim 10^4$ spinup. For the typical $\sim 10$~day periods of
main-sequence
stars \citep{mcquillan14}, minute-scale rotation periods would be
expected. However, as summarised below, typical observed WD rotation periods
are of order 1~day. Recent astroseismology of 300 red giants 
using {\it Kepler}
indicates that the degenerate stellar cores are slowly rotating already 
at the last stages of the red giant branch \citep{mosser12},
slower than currently explainable by stellar evolution models 
\citep{cantiello14}.     
An additional, later, source or sink
of angular momentum for WDs in close binaries could be
torques induced during common-envelope phases, Roche-lobe
overflow accretion
from a companion, or merger with another WD or stellar core
(e.g. \citealt{tout08}).

The theoretical possibility of fast WD rotation has been raised
recently
within the  ``spinup/spindown'' scenario
\citep{yoon04, distefano11, justham11, ilkov12}
of Type-Ia supernovae (SNe Ia), in which an accreting WD could
gain
enough angular momentum to be rotationally supported against
contraction and ignition as a SN Ia, beyond the Chandrasekhar
limit.
This would
also be a way of explaining some
``super-Chandrasekhar-mass'' SNe~Ia, in which the luminosity of
 the events suggests
an exploding mass significantly above the Chandrasekhar mass
(e.g. \citealt{howell06,silverman11,kamiya12}).
However,
to attain rotational support at masses significantly above
the non-rotating limit, WDs must have differential rotation, but
instabilities and magnetic torqueing might prevent this.
A handful of WDs for which rotation profiles have
been derived via astroseismology are indeed
consistent with solid-body rotation \citep{fontaine13}.
If spinup/spindown were a dominant SN Ia channel, the Galactic SN
Ia rate
per unit stellar mass,
the local stellar mass density, and  local density of WDs,
together
imply that $\sim 10^{-4} - 10^{-2}$
of the WD population would be fast rotators, with
periods of order 20~s (i.e. near breakup speed), for
spindown times of 0.1-10~Gyr, respectively \citep{maoz14}.
More generally, the existence or
absence of WDs even with moderately fast spin periods, of order
minutes, could
inform the question of whether sustained fast WD rotation is at
all possible.

Observationally,
 measurement of WD rotation is
difficult. The absorption lines
of hydrogen and helium in the spectra of the
most common DA- and DB-type WDs, respectively,
are highly pressure-broadened, but have a weak
narrower NLTE line core that is observable in H$\alpha$ and
H$\beta$.
Spectroscopic estimates of rotational broadening of the NLTE core
are quite uncertain, or give only
upper limits (see e.g. \citealt{karl05}, who obtain projected
rotational velocity estimates for about 20 DA WDs).
As detailed below, more precise WD rotation measurements have
 been mostly
limited to relatively rare types of WDs --- magnetic WDs,
by means of photometric variability, and WDs with optical-range metal lines,
based on spectroscopic observation of
their rotation-broadened NLTE metal-line cores.

Some 10-20\% of WDs have
surface magnetic fields $B\gtrsim 10$~kG, and about 5\% have
$B\gtrsim 1$~MG  \citep{kawka07,
holberg08, kepler13}. Magnetic fields can be
estimated
from Zeeman-splitting of
absorption-line cores, or by spectropolarimetric
measurements of circular polarization variations across line
profiles.
For strong magnetic fields, variations in the parallel
component
of the surface field can lead to
``magnetic dichroism'' --
a net circular polarization due to a dependence of
continuum opacity on field strength. The net
circular polarization, expressed by the ratio of Stokes
parameters, depends
on magnetic field roughly as $V/I\sim B/(10^9~{\rm G})$
\citep{angel81}.
Thus,
when $B\gtrsim 100$~MG, this effect can lead,  in
rotating WDs,
 to periodic photometric variations of order 10\%, (e.g.
\citealt{ferrario97}).

In the less strongly
magnetic WDs, localized
star spots may be able to form at sites where
the magnetic field inhibits convective motions,
resulting in lower effective temperatures there. Combined with
rotation,
this can
possibly lead to
photometric modulation.
Because the atmospheres of DA-type WDs
 are expected to
become fully radiative above  $T_{\rm eff}=$12,000--14,000 K,
and DB WDs above $T_{\rm eff}=$23,000--28,000 K,
hot WDs are not expected to have spots \citep{brinkworth05}.
\citet{brinkworth13} have
searched for photometric variability among 23 magnetic WDs.
Including some
previously known results, they obtain periods for 10 WDs, and
rough
possible period ranges for another 7, with  typical amplitudes
of 1--2\%.
They interpret these results
as rotation periods made observable by cool magnetic spots.

A second class of WDs permitting detection of
rotation are
DAZ- DBZ- and DZ-type, i.e. WDs displaying, in addition to
hydrogen or helium in their atmospheres,
strong metal lines in
their optical spectra
\citep[produced by pollution of the photospheres by accreted
dust
or rocky material;][]{farihi13}.
Metal-rich WDs
constitute $\sim$ 10--20\% of all WDs \citep{holberg08}.
Because of their higher atomic weight, the NLTE line
cores of the metal lines have smaller instrisic thermal widths
than those of the hydrogen lines, facilitating the measurement
of rotational broadening. Among 38 DAZ WDs,
\citet{berger05} measured
projected rotation speeds for 10 WDs, and
 upper limits of $v\sin i < 10$ \kms\ for the
other 28 objects.

Finally, WD rotation can be measured by means of astroseismology
of WDs
that have cooled, after a few 100 Myr, to effective temperatures
of
$T_{\rm eff}\sim$ 12,000 K, at which point
they undergo non-radial $g$-mode pulsations. The pulsations
typically
have periods of $\sim$ 100--1000 s, and up to
$\sim 1$~hr in very low-mass WDs
\citep{winget08}. WD rotation can induce splitting of
 pulsation frequencies in the power spectrum.
\citet{kawaler04} summarizes rotation periods obtained
 for 10 pulsating WDs in this way. A recent measurement of this
type
by \citet{greiss14}
using the {\it Kepler} Mission, for the DA WD KIC 11911480,
reveals a rotation period of 3.5~d.

Based on the various methods,
the successful detections of rotation, for $\approx 50$ WDs in
total,
indicate slow
rotations.
WD period distributions center on periods of order a day, with
few
periods of less than $\sim 30$~min [e.g. \citealt{kawaler04}; see
\citealt{ferrario97}, for an isolated (i.e. not in an interacting
binary)
 hot WD, likely a merger remnant, with a 12~min period],
but with many
lower limits suggesting significantly longer periods.
However, this is a small number of WDs,
and the studied stars are
not from among the most common WD types -- they are
either magnetic, with metal-rich atmospheres, or in the narrow
temperature
 range in which pulsations occur.
Rotation periods for larger samples of normal WDs would reveal
the
distribution of angular momentum  for the WD population as a
whole.

\subsection{Close companions to WDs}
\label{sec:intro_companions}
Substellar and stellar-remnant companions to WDs can not only
provide clues
to the final stages of stellar evolution, but may well play an
important
role in that evolution, e.g. by stripping stellar envelopes
during
common-envelope phases (e.g. \citealt{soker13}).
Brown-dwarf + WD systems
permit age-dating the brown dwarfs, and are important for understanding
their
 physics and demography (e.g. \citealt{debes11}).
Close double-WD systems are tracers of
stellar multiplicity (e.g. \citealt{badenes12}), prime candidates
for SN Ia
progenitors (e.g. \citealt{maoz14}),
and the main sources for future space-based
gravitational wave experiments (e.g. \citealt{amaro13}).
Planets in the habitable zones of WDs \citep{agol11}
may hold the best hopes for the
detection
of biomarkers \citep{loeb13}. The statistics of these populations
have begun to emerge, but much is yet to be learned.

A handful of brown-dwarf companions to WDs is known
(see, e.g. summary in
\citealt{debes11}), but it is clear that such systems are rare,
in analogy to the ``brown-dwarf desert'' of main-sequence stars.
Based on near-infrared excesses in  samples of WDs from SDSS,
\citet{steele11} and \citet{girven11} have estimated that $\sim 0.5$
 to 2 per cent of WDs have brown-dwarf companions.

About 50 close double WD systems are known, and statistical
studies
have estimated the fraction of such sytems among WDs at 2 to 20\%
\citep{maxted99}.
More recently, \citet{badenes12} found, among
4000 WD spectra
from SDSS, 15 WDs
with observed RV changes of $>250$~km~s$^{-1}$ between
consecutive
($\sim 15$~min)
exposures, strongly suggestive of double-WDs with separations
$a<0.05$~AU,
corresponding to periods of up to about 4 days, for typical WD
masses.
A statistical analysis interprets this result into
a $\sim$ 5--10\% true fraction of such
systems among WDs.

Turning to planets around WDs, there are no {\it bona fide} known
cases
of WDs with planets, but there is growing circumstantial evidence
for
their possible existence, despite theoretical questions if
planets can
survive engulfment by their parent stars during the giant stages
\citep[e.g.][]{villaver09, nordhaus13}.
For example, \citet{charpinet11} have used {\it Kepler}
to deduce the presence of
two Earth-sized planets orbiting a hot B subdwarf star (a red
giant
stripped of its envelope, and destined to become a WD), with
5.7~hr and
8.2~hr periods. Reflection or thermal re-emission of the WD
radiation
by the planets modulates the WD light by $\sim 50$~ppm.
Circumstellar disks of gas and dust are seen in some WDs,
by means of mid-infrared excesses (e.g. \citealt{hoard13}).
\citet{koester14} used Hubble Space Telescope
ultraviolet spectroscopy of 85 DA WDs to identify accreted
refractory
 elements in their atmospheres,
and deduce that 27\% to 50\% of the WDs are currently accreting
planetary debris. This material is predominantly of a rocky nature, 
occasionally even differentiated \citep{gaensicke12}.
In one WD, \citet{farihi13}
conclude from analysing the
abundance of oxygen in the WD atmosphere that the accreted
material consisted
of 26\% water, presumably from an asteroid-sized object that was
tidally
disrupted and accreted. \citet{debes12} argue that
the accreted
planetesimals
are perturbed into their close orbits via
resonance with a
Jupiter-mass planet.
In summary, the discovery of planets around WDs may be around the
corner.

\subsection{{\it Kepler} in quest of companions and rotation in a
small sample of WDs}

Below,
we search for possible signs of companions and rotation in a
sample of
14 normal WDs observed by {\it Kepler}. We detect and measure
periodicity
in at least five, and possibly up to seven of the 14 WDs, 
based on photometric variations much
smaller than
those detectable by previous WD variability studies.
We then discuss the possible physical origins of the observed
modulations.

\section{Sample Selection}
\label{sec:targets}

A sample of
14 WDs was observed with {\it Kepler} and studied by
\citet{ostensen10,ostensen11},
 as part of a program to search for pulsations among compact
stars
 (with a null result for pulsations among the WDs in the sample).
The sample is listed in
table 3 of \citet{ostensen11}, which is partly reproduced here
in Table~\ref{tab:main_tab}, including the WD
types, effective temperatures ($T_{\rm eff}$, to accuracies of
$\pm 500 K$),
and surface gravities ($\log g$, to accuracies of $\pm 0.3$),
as derived spectroscopically
by  \citet{ostensen11}. The gravities and temperatures, combined
with
models for WD radius vs. mass and temperature (e.g.
\citealt{wood95})
permit
estimating the WD mass to $\sim \pm 0.2$M$_\odot$.

Among these 14 WDs, one (WD1942+499 = KIC 11822535, a $T_{\rm eff}=36,000$~K
 DA WD)
was a previously known WD
in the {\it Kepler} field; eight WDs were discovered among
candidates selected
based on UV-excess in {\it Galex} photometry \citep{martin05};
two
were selected
based on UV excess in the SDSS-SEGUE survey (Yanny et al. 2009);
and seven WDs were
chosen from the USNO catalog based on the method of reduced
proper motions,
which detects relatively faint objects with large proper motions,
i.e.
nearby low-luminosity stars, primarily WDs. Several of the 14 WDs
were found
by more than one of these selection methods, and all 14
were spectroscopically confirmed as WDs and
 classified by \citet{ostensen10,ostensen11}

Significantly for our objective of possibly measuring rotation in
normal
WDs,
 there is no selection effect in the sample
that would obviously favour magnetic WDs. The fraction of magnetic
WDs
among the 14 is thus likely similar to the magnetic
fraction in the WD population as a whole, i.e. one or two WDs may
be
magnetic.
The sample selection methods do result, however,
in a hot WD sample, with all but one having $T_{\rm eff}>
14,000$~K, implying
cooling ages of order $10^8$~yr or less.
Twelve WDs are DA-type, one is a DB-type,
and one target (KIC~7129927) is likely
a double-degenerate system of
two DA WDs of comparable brightness, based the spectral
fitting of the Balmer-line profiles
\citet{ostensen11}.

 Apart from the ``long cadence'' ($\sim$30~min) data
available for all {\it Kepler} targets,
these 14 WDs also have ``short cadence'' ($\sim$1~min)
data, which permits searching for short periods.
Our analysis makes use of Public Releases 14--21 of
the long cadence data, and Public Release
21 of the short cadence data, which were downloaded from the
{\it Kepler} mission
archive\footnote{http://archive.stsci.edu/kepler}.
Table~\ref{tab:main_tab} includes the quarters
of data available for each WD, ranging from Q1--Q17.
Only a fraction of this time span was available to
\citet{ostensen10, ostensen11}
at the time of their study of this sample.

\begin{table*}
  \caption{White dwarfs}
  \label{tab:main_tab}
  \centering
  \begin{tabular}{p{1.cm}p{1.cm}p{1.cm}p{0.5cm}p{0.5cm}p{0.7cm}p{
.7cm}
p{.9cm}p{.9cm}p{2.6cm}p{1.4cm}}
  \hline
     KIC & RA & Dec & Kp & $T_{\rm eff}$ & $\log g$ &  Type &
Qtrs& Qtrs&Period &    $A$ \\
     &[deg] &[deg] &[mag]& [kK] & [cm/s$^2$] &&LC&SC& [days]  &
[ppm]   \\
     \hline
Periodic&&&&&&&&&&\\
     \hline
     5769827  & 283.688 & 41.088 & 16.6 & 66  & 8.2 & DA  &4,6&4
& 8.30 $\pm$ 0.04 & 818 $\pm$  44 \\
     6669882  & 283.942 & 42.118 & 17.9 & 30.5  & 7.4 & DA  &2 &2
&0.367 $\pm$ 0.009 & 821 $\pm$  16 \\
     6862653 & 291.692 & 42.327 & 18.2 & 16 & -- & DB & 2 &2 &
0.594  $\pm$ 0.02   & 507 $\pm$ 21 \\
     8682822 & 289.336 & 44.878 & 15.8 &  23.1 & 8.5  & DA
&5-9&1,5&  4.7 $\pm$ 0.3  & 60 $\pm$ 10 \\
     11337598 & 284.446 & 49.161 & 16.1 & 22.8 & 8.6 & DA
&3,10&3& 0.09328 $\pm$ 0.00003 & 311 $\pm$ 30 \\
     11514682 & 295.302 & 49.419 & 15.7 & 32.2  & 7.5 & DA
&3-14&2&  9.89 $\pm$ 0.06 & 62 $\pm$  6 \\
     11604781 & 288.537 & 49.611 & 16.7 & 9.1 & 8.3  & DA
&3,6,7&3& 4.89 $\pm$ 0.02 &  2060 $\pm$ 30 \\
     \hline
Non-periodic&&&&&&&&&&\\
     \hline

3427482 & 286.344 & 38.526 & 17.3 & --   & --  &DA&1&1&--  &$<600
$ \\
4829241 & 289.865 & 39.978 & 15.8 & 19.4 & 7.8 &DA&3-9&1,5&--&
$<50 $ \\
7129927 & 295.247 & 42.675 & 16.6 & --   & --  &DA+DA&3,5,6&3&--
&
$<80 $ \\
9139775 & 284.432 & 45.539 & 17.9 & 24.6 & 8.6 &DA&2&2&-- & $<700
$ \\
10198116& 287.497 & 47.286 & 16.4 & 14.2 & 7.9 &DA&4-6&4&-- &
$<70 $ \\
10420021& 297.311 & 47.579 & 16.2 & 16.2 & 7.8 &DA&5-10&2,5,6&--&
$<50 $ \\
11822535& 295.932 & 50.077 & 14.8 & 36.0 & 7.9 &DA&3-9&2,5-13&--
& $<20 $ \\

      \hline
     \\
  \end{tabular}
\end{table*}

\section{Time Series Analysis}
\label{sec:per_det}
\subsection{Method}

To search for periodic or quasiperiodic photometric modulation
in the WD light curves,  we calculated for each WD
 a fast Fourier transform (FFT) with zero padding,
and visually examined the  power spectra.
We did this for both the short- and long-cadence data,
searching for
a significant single peak in the
 power spectrum. When such a peak was detected,
  the most likely period was determined from a Gaussian fit to
the power
spectrum peak.
 The amplitude of variability was
 determined by fitting a linear combination of a sine and a
 constant to the light curve, at the period detected.
Uncertainties
in the period and the amplitude were estimated by varying these
parameters
until $\Delta \chi^2=1$ was obtained.

The detection threshold for periodicity varies from object to
object, depending
on its brightness, the number of quarters of observation, and
possible noise
characteristics specific to the object (e.g. due to backgrounds
from stars at
small separation). To determine these thresholds we planted, in
all WDs
without detected periods, sine functions with a
period of 0.5~d (which is characteristic of those WDs with a
detected signal),
with a range of amplitudes, and found the minimum variation
amplitude that
is discernible in a visual examination of the power spectrum.
These limits
are also listed in Table~\ref{tab:main_tab}.

To search for transient or quasiperiodic modulation, as produced
by evolving
spots on normal stars (see McQuillan et al. 2014), we also
applied an
autocorrelation-function (ACF) analysis to all of the WD time
series. We did
not find significant results. This, and the power-spectrum
analysis results below,
indicate a stable modulation.

\subsection{Results}
\label{sec:results}
Among the 14 WDs in the sample, we detect a periodic signal in
at least five, and possibly up to seven WDs. 
All seven detections are in the long-cadence data, with periods in the 
range of 2~hr to 10 days, and semi-amplitudes of 60--2100~ppm.
In the
short-cadence data, we rediscover the same long-timescale 
periodicities that we detect in the long cadence data, but
we do not detect any short-timescale periodic modulations, 
other than several artificial periods already noted by
\citet{ostensen10,ostensen11}.
In five cases, the detection is unequivocal. In one case, KIC~11514682
the semi-amplitude (62 ppm) is low, 
yet we detect the same period in the data for
each of three separate years (with four quarters each), lending confidence
to its reality. For another WD, KIC~8682822, which has the lowest 
variation amplitude (60 ppm),
we have only two years (eight quarters) of data. Although the same 
amplitude-spectrum peak frequency appears in both years, it sits on top of
a broad low-frequency bump, which could be due to noise, aliasing of
the real period, or real quasi-periodicity of the signal. We have further 
tested the reality of these two detections using a Monte-Carlo
simulation. In each case, we have produced 1000 different
time-scrambled versions of the light curve, and calculated the Fourier amplitude
spectrum for each version. In none of the simulated cases was the amplitude of the highest
peak in the spectrum, at any frequency, as high as the observed peaks. 
Nevertheless, we consider the
detection of periodicity in KIC~11514682 as not completely conclusive, 
and in KIC~8682822 as only tentative.
 
\begin{figure}
 \centering
 \includegraphics[width=\linewidth]{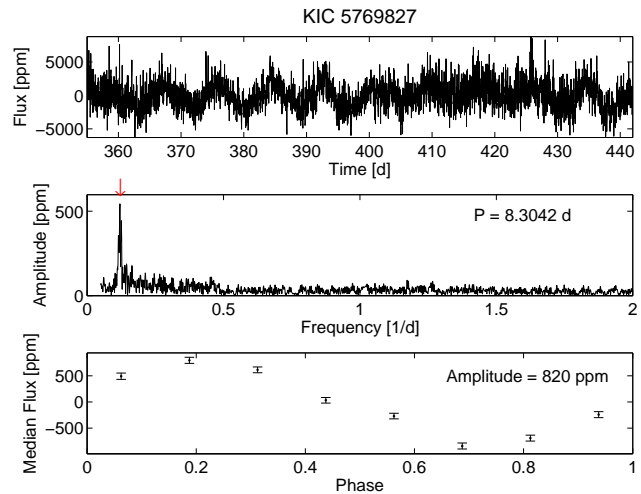}
\caption{KIC\,5769827 -- Top panel: segment of the normalised
light curve; middle panel: FFT amplitude spectrum; bottom panel:
phase-folded and binned light curve.}
\label{fig:5769827}
\end{figure}

\begin{figure}
  \centering
 \includegraphics[width=\linewidth]{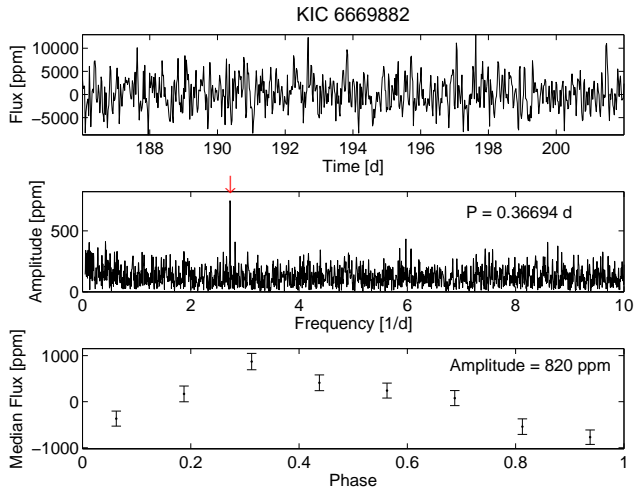}
  \caption{Same as Figure~\ref{fig:5769827}, for KIC\,6669882.}
  \label{fig:6669882}
\end{figure}

\begin{figure}
  \centering
 \includegraphics[width=\linewidth]{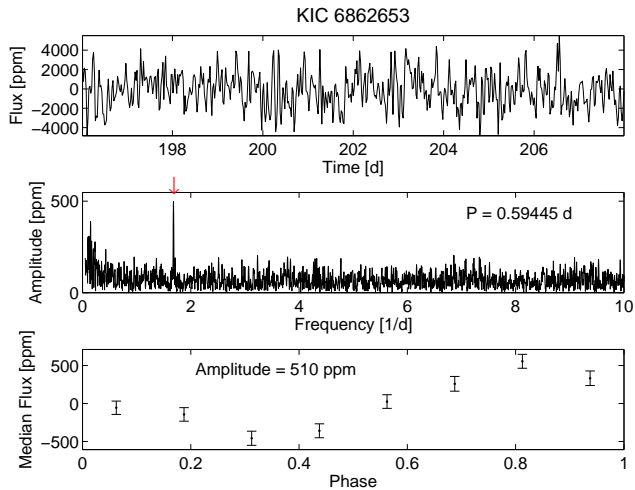}
  \caption{Same as Figure~\ref{fig:5769827}, for KIC\,6862653.}
  \label{fig:6862653}
\end{figure}

\begin{figure}
  \centering
  \includegraphics[width=\linewidth]{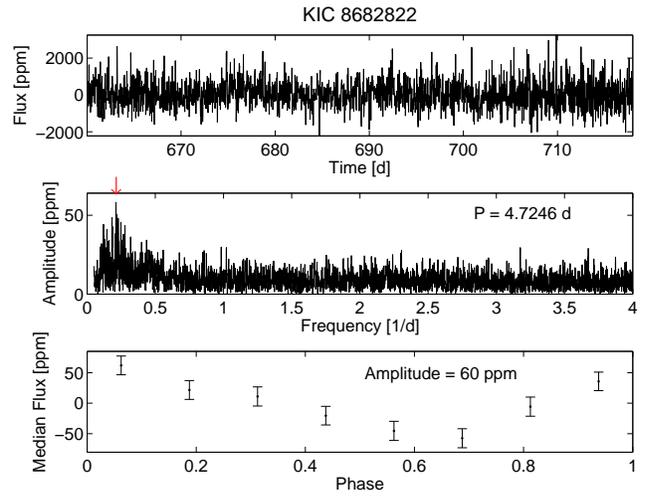}
  \caption{Same as Figure~\ref{fig:5769827} for KIC\,8682822.}
  \label{fig:8682822}
\end{figure}

\begin{figure}
  \centering
  \includegraphics[width=\linewidth]{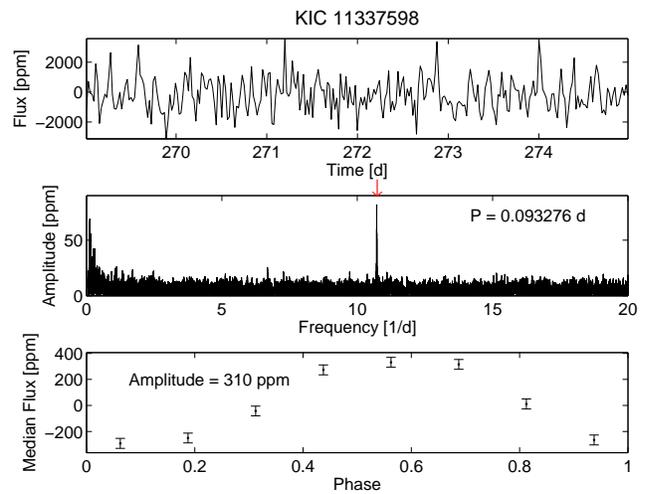}
  \caption{Same as Figure~\ref{fig:5769827}, for KIC\,11337598.}
  \label{fig:11337598}
\end{figure}

\begin{figure}
 \centering
 \includegraphics[width=\linewidth]{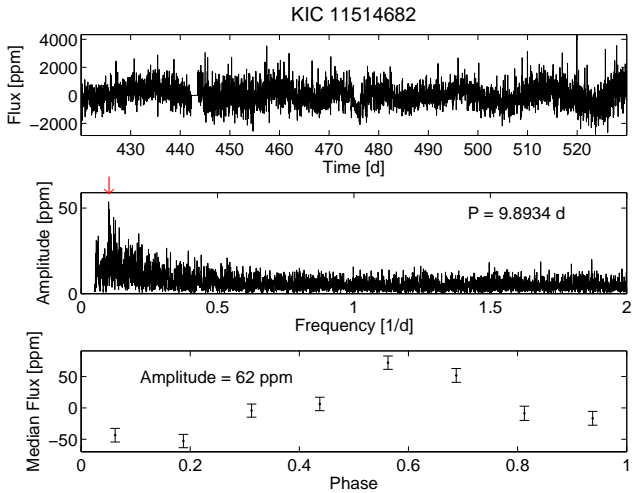}
 \caption{Same as Figure~\ref{fig:5769827}, for KIC\,11514682.}
 \label{fig:11514682}
\end{figure}

\begin{figure}
  \centering
  \includegraphics[width=\linewidth]{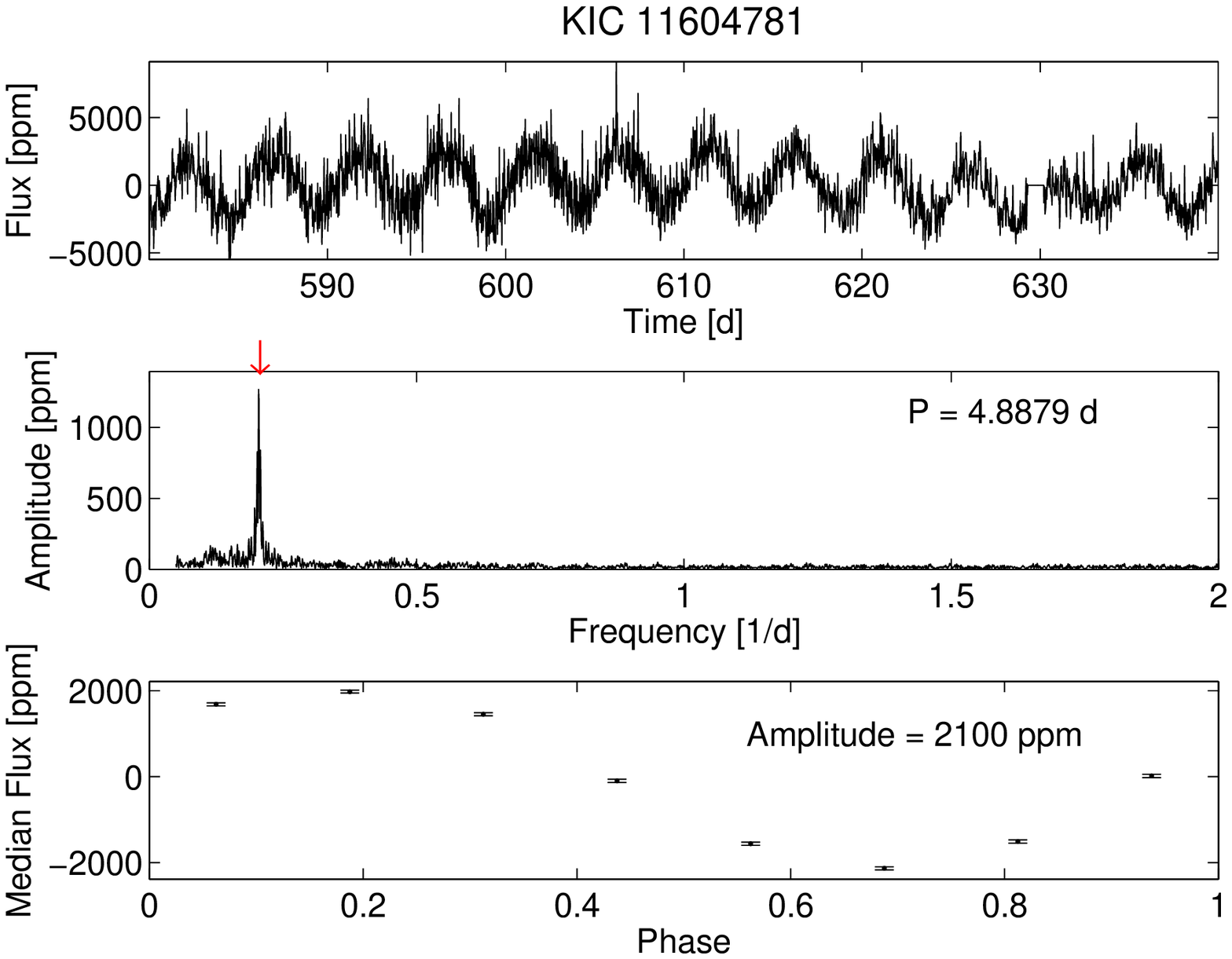}
  \caption{Same as Figure~\ref{fig:5769827}, for KIC\,11604781.}
  \label{fig:11604781}
\end{figure}

The periods,
semi-amplitudes, and their uncertainties, are
listed in Table~\ref{tab:main_tab}.
Figures~\ref{fig:5769827}--\ref{fig:11604781} show, for each of
these seven WDs, a section of
the processed light curve,
the amplitude spectrum of the whole data set, and the period-folded
light curve, binned into eight bins. For each bin we give the
median of the points in that bin and draw an error bar, estimated
as 1.48 times the median absolute deviation (MAD)
around the median, divided by $\sqrt{n_{bin}}$,
where $n_{bin}$ is the number of points in that bin. In all seven cases,
the light curves appear sinusoidal in shape.

\section{Physical mechanisms for periodic photometric
variability}
The possible interpretations of photometric modulation are
 not always
unique.
We therefore begin by discussing, in general,
 the possible sources of periodic photometric
variability in the range observed for these WDs.

\subsection{Transits by a companion object}
A companion object in a high-inclination binary orbit
can produce periodic variability by means of
transits.
 Rocky orbiting objects of sizes $\sim 50-200$~km  could produce
the observed small amplitudes.
However, any transiting object would result
in a characteristic dip, visible in
the folded light curve during only a small fraction of an orbital
period, and the power spectrum would display a series of peaks at harmonics of 
the orbital period.
The folded light curves for the WDs with periodicities are not of
this type,
and thus transits can generally be ruled out. We note that, for all 
the other mechanisms that we discuss below, and assuming circular orbits,
one expects roughly sinusoidal light curves, as observed.

\subsection{Beaming due to reflex motion induced by a companion}
Another potential cause of periodic
variability is special-relativistic beaming of the WD light,
 due to orbital motion of the WD, induced by a companion.
In our sample, the WDs are hot, with the
 {\it Kepler} bandpass always well on the
Rayleigh-Jeans side of the thermal spectrum.
The periodic signal that 
beaming will produce, for this case, has relative
semi-amplitude
$A\sim (v \sin i)/c$,
where $v$ is the orbital velocity
of the WD,
and $i$ is the inclination angle between the orbital plane's axis and the 
line of sight
\citep{loeb03,zucker07,faigler11,mazeh12}.
For a WD mass
of $M_{\rm wd}$ (typically $0.6$~M$_\odot$) and a companion mass $M_2$,
an observed period $P$ implies a ``mass function''
$$
\frac{M_2 ~\sin i}{M_\odot}
\left(\frac{M_2+M_{\rm wd}}{1.6 M_\odot}\right)^{-2/3}
$$
\begin{equation}
=    \left(\frac{A}{520~ {\rm ppm}}\right)
\left(\frac{P}{1~{\rm d}}\right)^{1/3}.
\end{equation}

\subsection{Reflection from a companion}
Alternatively to beaming, periodic modulation can be produced by
WD light
that is
reflected from an orbiting cool companion, as it goes through
illumination
phases. The modulation's semi-amplitude in this case is $A\sim
(D\sin
i/8)(R_2/a)^2$, where $D$ is the albedo of the companion, $R_2$
is its radius and $a$ is the separation.
For  a $0.6~$M$_\odot$ WD and a companion of significantly lower
mass, the minimum
companion radius required to explain the observed modulation
(obtained when
both $D=1$ and $\sin i =1$) is then
\begin{equation}
R_2 > 0.1~R_\odot \left(\frac{A}{100~{\rm ppm}}\right)^{1/2}
\left(\frac{P}{1~{\rm d}}\right)^{2/3}
\left(\frac{M_{\rm wd}}{0.6~M_\odot}\right)^{1/3}
\end{equation}
Companions with 
radii up to $R_2\lesssim 0.2~R_\odot$ could be
planets (Jupiter-mass planets up to such radii have been discovered,
see e.g. \citealt{baraffe14})
or brown dwarfs.
The brown-dwarf reflection option, however,
is unlikely to be relevant for more than one or two of the 
WDs in the sample, because there is a $\lesssim 2$\% fraction of
brown-dwarf companions to WDs \citep{steele11,girven11}.
A minimum companion radius above $R_2\sim0.2 R_\odot$ 
implies a
hydrogen-burning
star with $T_{\rm eff}\gtrsim 2800$~K (e.g. Hillebrand \& White
2004),
which can be tested via the
presence of a thermal IR signal (see below).

\subsection{Thermal emission from a companion}
For tight WD-companion orbits,
 a sub-stellar companion can be heated non-uniformly (e.g. on its
tidally locked
day side) by the WD. The thermal re-rediation from the heated
companion
hemisphere, as it goes through apparent phases, may produce 
observed
photometric modulation.
The companion's equilibrium insolation temperature will be
\begin{equation}
T_{\rm 2}= T_{\rm wd} \left(\frac{1-D}{f}\right)^{1/4}
\left(\frac{R_{\rm wd}}{a}\right)^{1/2},
\end{equation}
where $f$ is a factor between 4 and 2,
depending on whether
the companion redistributes absorbed heat over its entire
surface (in which case there will be no modulation from thermal
re-radiation)
or over only the hemisphere facing the WD (in which case it is
the temperature only of that hemisphere). If heat
redistribution is particularly inefficient, then a single
effective temperature for the companion is poorly defined.

The flux ratio between the WD and the companion at a
wavelength $\lambda$ is
\begin{equation}
\label{eq:bbratio}
\frac{f_{\rm WD}}{f_{\rm 2}}=
\frac{B_{\lambda}(T_{\rm
wd})}{B_{\lambda}(T_{2})}\left(\frac{R_{\rm
wd}}{R_2}\right)^2,
\end{equation}
where
$B_{\lambda}(T_{\rm eff})$ is the thermal
flux per unit wavelength at wavelength $\lambda$
from an object of effective temperature
$T_{\rm eff}$, $T_{\rm wd}$ and $T_2$ are the effective
temperatures of the WD and the companion, respectively,
and $R_{\rm wd}$ and $R_2$ are their respective radii.
The Wien tail
from the thermal emission  of an irradiated companion can enter
the {\it Kepler} photometric bandpass, which extends from $\sim
4400-8800$~\AA.
Multiplying the numerator and denominator in
the ratio in Eq.~\ref{eq:bbratio} by the throughput
and integrating each of them
over the bandpass gives twice the maximum modulation
semi-amplitude
(corresponding to
a high line-of-sight inclination of the orbital plane) via this
mechanism.

In addition, using Eq.~\ref{eq:bbratio},
the time-independent signature
 of a possible stellar companion with known effective temperature
of its own, whose mass is indicated, e.g., by beaming,
can be tested by means of the companion's expected near-infrared
or optical thermal emission, which may be comparable to that from
the WD,
or even dominant.

\subsection{Rotation plus cold magnetic starspots}
As noted, periodic photometric variations can be caused by WD
rotation,
combined with a spotted surface. In analogy to normal stars,
stars spots in WDs may be able to form at sites where
the (possibly weak) magnetic field  inhibits convective motions,
resulting in lower effective temperatures there. Spots are not
expected
in WDs whose atmospheres are fully radiative ---
DA WDs with $T_{\rm eff} \gtrsim$ 14,000 K,
and DB WDs with $T_{\rm eff}\gtrsim$ 28,000 K. Only two of the
variable WDs
in our sample are cool enough for spots to explain the observed
variations.

\subsection{Rotation plus magnetic dichroism}
Periodic photometric modulation can be caused by a heterogeneous
atmospheric
continuum opacity due to magnetic dichroism. As noted in 
Section~\ref{sec:wdrotation}, the
amplitude
depends linearly on magnetic field as  $\sim B/(10^9~{\rm G})$.
The observed WD
semi-amplitudes listed in Table~\ref{tab:main_tab}, of $A\sim
10^{-4}$--$10^{-3}$,
 would then imply surface fields of $B\sim 100$~kG to 1~MG. As we
do not know
the magnetic field strengths for this WD sample, each variable
WD, individually,
could be explained with this mechanism. However, only 5--10\% of
the general WD
population are this highly magnetic, and therefore from a
statistical point of
view it is unlikely that, from our sample of 14 WDs, more
than one or
two would be magnetic. In any event, this possibility can be
tested in the
future by
measuring the magnetic fields of the seven periodic WDs.

\subsection{Rotation plus hot spots from magnetically channelled
ISM accretion}
Another rotation-based mechanism
for photometric periodicity is
 accretion by the WD of gas from the interstellar medium
(ISM, e.g. \citealt{liebert80}),
and channeling of the material onto the WD's magnetic poles,
producing
hot spots that cause the observed modulation.

A WD moving through the ISM
at velocity $v_{\rm ran}$ will have its geometric cross-section,
 $\sigma_{\rm g}=\pi R_{\rm WD}^2$,
effectively
increased through gravitational focusing to
 $\sigma_{\rm gf}=\sigma_{\rm g}(1+v_{\rm esc}^2/v_{\rm ran}^2)$,
where $v_{\rm esc}$ is the escape velocity from the WD surface,
which is
$\sim 4500$~km~s$^{-1}$ for a typical WD of mass 
$M_{\rm wd}=0.6~$M$_\odot$
and radius
$R_{\rm WD}\sim 8000$~km. The minimum
mass accretion rate through such direct impact of ISM gas on the
WD is thus
\begin{equation}
\dot M_{\rm gf}=\rho_{\rm ism}~ \sigma_{\rm gf}~v_{\rm ran}
\end{equation}
$$
\approx 1.6\times 10^5
~{\rm g~s}^{-1} \left(\frac{n_{\rm ism}}{1~ {\rm
cm}^{-3}}\right)\left(\frac{M}{0.6~{\rm
M}_\odot}\right)\left(\frac{v_{\rm ran}}{35~
{\rm km~s}^{-1}}\right)^{-1} ,
$$
where $\rho_{\rm ism}$ and $n_{\rm ism}$ are the ISM mass and
volume densities,
respectively, and the fiducial $v_{\rm ran}=35$~km~s$^{-1}$ is
the 1$\sigma$
random velocity dispersion of disk stars from \citet{Han95}.

However, a potentially much larger accretion rate can arise in
the
context of Bondi-Hoyle-Lyttleton (BHL) accretion
(\citealt{hoyle41,
bondi44}; see \citealt{edgar04} for a review),
in which
the gravity of the moving object
focuses material from a much greater volume into
a hydrodynamical downstream conical wake, that the
moving object then  accretes.
The BHL accretion rate is
\begin{equation}
\dot M_{\rm BHL}=4\pi G^2 M^2 \rho_{\rm ism} v_{\rm ran}^{-3}
\end{equation}
$$
=2.7\times 10^9  ~{\rm g~s}^{-1} \left(\frac{M}{0.6~{\rm
M}_\odot}\right)^2
\left(\frac{n_{\rm ism}}{1~ {\rm cm}^{-3}}\right)
\left(\frac{v_{\rm ran}}{35~
{\rm km~s}^{-1}}\right)^{-3}.
$$
%
%
The gravitational energy of
this mass flow will be thermalized on
impact on the surface of the WD,
giving the WD an increment to its bolometric luminosity,
$L_{\rm bol}$,
 of
\begin{equation}
\Delta L_{\rm acc}=\frac{1}{2}\dot M_{\rm BHL} v_{\rm esc}^2
\end{equation}
$$
\approx 3\times 10^{26}~  {\rm erg~s}^{-1}
\left(\frac{M}{0.6~{\rm M}_\odot}\right)^{10/3}
\left(\frac{n_{\rm ism}}{1~ {\rm cm}^{-3}}\right)
\left(\frac{v_{\rm ran}}{35~
{\rm km~s}^{-1}}\right)^{-3},
$$
where we have assumed that $R_{\rm wd}$ scales roughly as
$M^{-1/3}$.

Supposing the magnetic field around the WD channels this gas to
the magnetic
poles of the WD, and the magnetic axis is misaligned with the
rotation axis,
then rotating hot spots could form, that could produce
photometric variability.
In ``polar'' cataclysmic variables (CVs), an accretion disk
around a WD is
disrupted within a radius where the gas pressure is less than
 the magnetic energy density, and the mass flow is channeled to
the magnetic
poles (e.g. \citealt{cropper90}).
Polar CVs are seen to occur only in WDs with $B\gtrsim
10^5$~G. However,
the accretion rates in CVs are $\sim 10^7$ times higher than the
BHL
 accretion rates considered here, so
the gas density of the BHL accretion flow (which may make a transition
to an accretion disk geometry at a small distance from the WD)
is also likely orders
of magnitude smaller than the density in the accretion disk of a
CV.
Channeling of the accreted material
to the poles may then be possible, even with the kG-or-less
magnetic fields of common WDs.

The accreted power will result in a temperature increment,
$\Delta T$,
within the hot spot, which will translate to a relative increment
in the
optical luminosity of the WD,
\begin{equation}
A=\left(\frac{\Delta L}{2L}\right)_{\rm opt}=\frac{\Delta L_{\rm
acc}}{8 L_{\rm bol}}
 \end{equation}
$$
=100~{\rm ppm}~
\left(\frac{\Delta L_{\rm acc}}{10^{28}~  {\rm
erg~s}^{-1}}\right)
\left(\frac{T_{\rm eff}}{13~{\rm kK}}\right)^{-4}
\left(\frac{M_{\rm wd}}{0.6~M_\odot}\right)^{2/3} .
$$
The incremental accretion power in the polar hot spots thus
could, in principle,
produce enough contrast to give the observed relative variation
amplitudes of
$A\sim 10^{-4}$--$10^{-3}$, if the WD mass, the ISM density, and
the WD velocity
result in a high enough accretion rate, and the WD temperature is
not too high
(which produces a high WD luminosity and hence a lower contrast
of the hot
spots).

\citet{koester06} considered BHL accretion from the ISM
onto WDs, by
way of explaining the metals observed in DAZ WD atmospheres, as resulting
from ISM
solar-abundance gas
accretion. DAZ WDs would then be those WDs, among the full WD
population,
 with the highest ISM accretion
rates. From their calculated
diffusion
times of metal atoms in WD atmospheres, and the observed calcium line
strengths in a sample of 38 DAZ
 WDs,
they derived accretion rates for these WDs. Assuming the accretion
is from ISM gas with a solar abundance, they obtained total ISM
mass
accretion rates for DAZ WDs in the range
$\dot M\sim 10^8 - 10^{11}{\rm g~s}^{-1}$, with a median rate of
  $\dot M\sim 5\times 10^9~{\rm g~s}^{-1}$. As they
knew $M_{\rm wd}$ and $v_{\rm ran}$ for every WD in their sample, they
could also
derive $n_{\rm ism}$ for each WD, under
the assumnption of BHL accretion. \citet{koester06}
 concluded that most of their DAZ
WDs were passing through warm, partly-ionized,
ISM clouds with $n_{\rm ism}\sim 0.01 - 1$~cm$^{-3}$.

However, by now it has become quite clear that DAZ WDs
accrete their metals from circumstellar disks of rocky debris,
rather than from
the ISM (e.g. \citealt{farihi10, farihi12}).
Recalling that, in the ISM accretion picture,
the 10--20\% of WDs with metal lines are supposedly those WDs
with the highest
accretion rates, the rates found by \citet{koester06}
would
then constitute
firm upper limits on any ISM accretion. On the other hand, if
gas accreted from the ISM is channeled to the magnetic poles, it
is conceivable
that the ISM metal atoms
would remain confined to the polar regions, before sinking
locally into the WD, and hence an ISM-accreting
WD might not appear as a DAZ WD after all. In such a scenario,
one may expect
to see an accreting WD transitioning spectroscopically, 
over a rotation period,
between
a DA and DAZ type. Indeed, WD rotation coupled with surface chemical 
inhomogeneity, resulting from anisotropic accretion, has been proposed
as another, related mechanism to explain WD periodicity 
(e.g. \citealt{holberg11}). We conclude that our proposed
scenario of rotation combined with
ISM accretion onto magnetic hot spots may be viable for
explaining
the observed periodicities and amplitudes, but, as we show below,
rather extreme conditions are required in order to explain some
of the observed variation amplitudes.

\subsection{Rotation plus non-uniform UV line opacity and optical fluorescence}
We consider one last possible mechanism to explain the
periodic variability seen in our WD sample, based on rotation combined
with UV opacity. As noted above, 
\citet{koester14} found, using UV spectroscopy with HST, that 56\% of WDs,
in the temperature range 17~kK to 27~kK (similar to the temperatures of 
our sample WDs), display metal absorption
 lines, presumably the result of recent or ongoing accretion of rocky circumstellar 
material. Examination of the spectra suggests that of order 1\% of the UV flux 
is absorbed in these lines. Some fraction of the energy of the 
absorbed photons will be ``degraded'' and re-emitted as a pseudo-continuum 
of optical and IR 
photons, a process known as fluorescence. Calculations of fluorescence for
 the case of Type-Ia supernovae by \citet{pinto00} (in which there is extremely heavy line-blanketing by metal species) show that the fractional flux
absorbed in the UV and re-emitted in the optical is comparable. If this effect
operates in WD atmospheres as well, and if the accretion onto the WD is not 
fully uniform (e.g. due to some 
channeling of the flow to the poles by weak magnetic 
fields), a non-uniform WD optical surface brightness could result. For example,
if the UV opacity, averaged over a WD hemisphere, differs by 1\% between 
two hemispheres, and 1\% of the UV flux is absorbed and fluoresces in the 
optical, WD rotation could then lead to the $A\sim 10^{-4}$ modulations
that we see. Remarkably, the incidence of UV opacity found by \citet{koester14},
and the incidence of WD variability that we find, are both 50\%. Thus, this
mechanism is one (and the only one among all of the mechanisms above) that could explain 
all of the cases of variability among WDs in our sample. An extreme example
of a WD with non-uniform UV opacity and rotation, leading to periodic 
variability is the isolated WD GD 394, which is extremely metal polluted
and shows EUV variations with 25 per cent amplitude and a 1.15~d period
\citep{dupuis00,chayer00}.
However, the 
viability of UV-to-optical fluorescence for the case of mildly UV-opaque 
WD atmospheres 
needs to be 
demonstrated by detailed calculations. Observationally, UV spectroscopy of our 
full WD sample could test whether there is a correspondence between the 
presence of UV absorption lines and optical variability.

\begin{table*}
  \caption{Possible interpretations }
  \label{tab:interpretation}
  \centering
\begin{tabular}{r c c c c c c}
  \hline
     KIC \,\,\, & cool  & magnetic & accretion& UV-optical &beaming&reflection/\\
            & spot      & dichroism&spot      &fluorescence & +WD/BD &re-radiation\\
     \hline
     5769827  &  -- &?  &? &+&--&--\\
     6669882  & --  &?  &? &+& ?&+\\
     6862653 & +    &?  &+ &+& ?&+\\
     8682822 & --   & + &+ &+& ?&--\\
     11337598 &--   & ? &? &+& ?&+ \\
     11514682 & --  & + &? &+& ?&--\\
     11604781 &+    & ? &+ &+&--&--\\
     \hline

  \end{tabular}

Notes: +, possible; ?, possible but statistically unlikely; --, excluded.
\end{table*}

\section{Interpretations for individual objects}
\label{sec:indiv}
We consider now the possible physical interpretations for each of the
seven WDs.
Table~\ref{tab:interpretation} summarises how well the models
fare for
each WD. UV line absorption plus optical fluorescence and WD rotation,
if it is a physically viable mechanism, 
could explain all of the variable cases, and 
hence we do not repeatedly state it for each individual WD below.  

\subsection{KIC~5769827 = J18547+4106}
Based on the WD surface gravity, the temperature,
and their uncertainties, as
found by \citet{ostensen11}, this DA WD
has a mass in the range $M\approx 0.6-1.0~$M$_\odot$. Its
photometry reveals a
8.3~d period with a semi-amplitude of 820 ppm.
The effective temperature is too high for cool magnetic spots.
The variations can result from rotation combined with
 magnetic dichroism if $B\sim 1$~MG, which is possible, but
{\it a priori} unlikely.
For the observed amplitude to be attributed to
ISM accretion plus magnetic channeling, the accretion luminosity
would
need to be $\gtrsim 6\times 10^{31}$~erg~s$^{-1}$, which is
extreme,
but possible, if the WD mass
is in the high end, the ISM density is quite high, and $v_{\rm
ran}$ is
rather low.
For beaming to explain the period and amplitude,
the unseen companion must have a mass $M_2> 15$~M$_\odot$, i.e.
only a massive
black hole could do it, and therefore it is very unlikely.

If variations are
due to reflection from a companion, the companion
radius would need to be at least $R_2> 1.2 R_\odot$, i.e the
radius of a
G star or more massive, whose luminosity would completely
dominate
over the WD in the optical. Rotation plus UV-to-optical line fluorescence, 
rotation plus magnetic dichroism
(implying a high magnetic field), or rotation plus a high ISM
accretion rate
thus appear to be the only possibilities
in this case.

\subsection{KIC~6669882 = J18557+4207} This DA WD, which
has a mass in the range $M\approx 0.40-0.55~$M$_\odot$ has
a 820 ppm semi-amplitude modulation with a period of 8.1~hr.
\citet{ostensen11} report a 127~s
periodicity in the short-cadence data, indentified by us as well,
which they conclude is an artifact.
As in KIC~5769827, above,
the WD atmosphere is too hot for the formation of convective
spots.
Rotation with magnetic dichroism would again require $B\sim
1$~MG.
Rotation with ISM accretion onto magnetic hot spots is likewise
possible,
but requires a high accretion rate, $\sim 2\times
10^{30}$~erg~s$^{-1}$,
difficult to achieve with the given low WD mass, unless $v_{\rm
ran}$ is very
small.
The observed periodicity
and photometric amplitude could be induced by beaming due to a
$M_2 > 1$~M$_\odot$ companion. This possibility can be ruled out
for
a main-sequence companion, as a
G-star would completely dominate
 over the WD
in the optical spectrum. However,
beaming cannot be presently ruled out if the companion is another
WD (though a quite massive one, especially considering the $\sin i$ factor).
In such a case,
radial velocity measurements using the thermal cores
of the Balmer lines will easily detect oscillations of
semi-amplitude $K\approx 200$~km~s$^{-1}$.
However, {\it a priori} the probability of detecting a companion
 WD that induces such RV variations is $\lesssim 1\%$
\citep{badenes12}.

Alternatively, the observed modulation could also be caused by
reflection
off a cool
companion with a minimum radius $R_2> 0.13~R_\odot$, i.e. a
Jupiter-like planet 
or a brown dwarf.
The brown-dwarf
option, as noted above, is {\it apriori} unlikely, because of the rarity of
brown-dwarf
companions to WDs \citep{steele11,girven11}.
The insolation temperature of the companion side facing the WD,
assuming
it is tidally locked in this configuration, would be
1500--2300~K for a range of albedo values $D=0.9$ to 0, making
this a ``hot''
companion. For a Jupiter-sized planet with its
day side heated to $>2000$~K, the thermal re-emission by the planet,
combined with the apparent
orbital phases,
could also give the observed amplitude of modulation, making thermal
re-radiation
by a substellar companion another option.
Thus, the possible explanations for the observed modulation are
rotation plus magnetic dichroism (though unlikely), ISM accretion
(questionable),
beaming due to the presence of
a massive cool WD companion (also unlikely),
and reflection, or re-radiation, from a giant-planet or
brown-dwarf
companion.

\subsection{KIC~6862653 = J19267+4219}
This is a $T_{\rm eff}\sim 16,000$~K, DB-type, WD, with 14.3~hr
period modulation
of 510 ppm semi-amplitude. \citet{ostensen11} could not
model well its spectrum, and hence its surface gravity is
unknown, and we
will assume a typical WD mass of $0.6~$M$_\odot$.
In DB WDs, atmospheres are convective
up to $T_{\rm eff}\gtrsim 23,000$~K, and hence cool magnetic
star spots are theoretically
possible in this case, and could combine with rotation to cause
 the observed photometric modulation. Alternatively, rotation
combined with magnetic dichroism
would imply a surface field of $B\sim 500$~kG, for which the
probability is
$\lesssim 5$\%. Rotation plus magnetic polar hotspots from ISM
accretion is,
again, another
option, if $\Delta L_{\rm acc}\sim 1.5\times
10^{29}$~erg~s$^{-1}$,
which is
possible
for some combinations of $M$, $n_{\rm ism}$, and $v_{\rm ran}$.
A $M_2> 0.9$~M$_\odot$ companion
in orbit with the WD with such a period could
produce this amplitude of variability by means of beaming,
but would again dominate the optical flux if it were a
main-sequence star,
and thus only beaming due to a cool
 WD companion is possible, though
 unlikely, because of the rarity of double WDs
with such close orbits \citep{badenes12}. This possibility
can be tested by the expected large $K\approx 200$~km~s$^{-1}$
radial
velocity variations.
Reflection of the WD light
from a $R_2> 0.15~R_\odot$ sub-stellar object could also explain
the
observations. It is {\it a priori}
 unlikely to be a brown-dwarf, leaving the
possibility of a giant planet. Its
insolation
temperature would be $\sim 500-900$~K.
In summary, in this case the most likely possibilities are cool
magnetic spots,
ISM-accretion-induced hot spots, and reflection from a planet.

\subsection{KIC~8682822 = J19173+4452}  This is a
DA WD of mass $\sim 0.8-1.2 M_\odot$, in which we
detect (tentatively, as explained in Section~\ref{sec:results})
a very small-amplitude  modulation of
60~ppm, with a  period of 4.7~d.
As in most of the WDs in the sample,
the temperature is too high for spottedness,
but rotation could work if combined
with  dichroism from a $B\sim 50$~kG field
(which is not exceedingly rare), or with an ISM accretion rate
of $\Delta L_{\rm acc}\sim 8\times 10^{28}$~erg~s$^{-1}$,
(which is plausible
given the large WD mass).
 If orbital beaming is the explanation,
the companion would have a mass of at
least $M_2> 0.1~$~M$_\odot$, if the WD is of mass $M_{\rm wd}>
0.8~$M$_\odot$,
 i.e. $M_2$ would be above the hydrogen-burning limit for stars.
To be subdominant at $2.2\mu$m (e.g. in the 2MASS survey),
the companion would need to have a temperature
$T_{2}\lesssim 2000$~K, but hydrogen-burning stars have
$T_{\rm eff}\gtrsim 2800$~K. However,
another WD that is cool and of very low mass, or cool and in a
low-inclination
(nearly face-on)
orbit could induce beaming at a level consistent with the
observed variations.
The expected RV semi-amplitude of the visible WD would be
$K=15$~km~s$^{-1}$, which
is challenging but observable.
If the source of photometric variability is, instead,
 reflection off a cool companion, it implies
a minimum companion radius $R_2\gtrsim  0.2~R_\odot$,
or even larger if
the albedo is not unrealistically
close to $D=1$, and the inclination is not close to $90^{\circ}$;
this
required radius is at the limit of known
substellar
objects (e.g. \citealt{baraffe14}),
and hence the thermal emission from such a hydrogen-burning
companion would
again have been detected.
Thus, the possible (though not equally plausible) options are
WD rotation plus dichroism or ISM accretion,
or beaming due to a cool WD companion that is either of low mass
or
in a near-face-on orbit.

\subsection{KIC~11337598 = J18577+4909} This $\sim
0.8-1.2$~M$_\odot$
DA WD has the shortest
periodicity that we find, $P=2.24$~hr, with a semi-amplitude of
310 ppm,
already noted by  \citet{ostensen11} and identified by them as
a possible rotation signature.  $T_{\rm eff}$ is again
 too high for spots.
Rotation combined with magnetic
dichroism from a $B\sim 350$~kG field,
or ISM accretion onto magnetic-pole hot spots with
$\Delta L_{\rm acc}\sim 6\times 10^{29}$~erg~s$^{-1}$,
could be the explanation. If the WD mass is indeed high,
such an accretion rate may be achievable.
If beaming due to a companion were to produce the modulation,
it would imply a minimum companion mass of
$M_2>0.25$~ M$_\odot$.
To avoid
being dominant at $2.2\mu$m, a main-sequence
 companion would need to have a temperature
$T_{\rm eff, 2}\lesssim 2000$~K, considerably
 less than the $T_{\rm eff}\gtrsim 3200$~K of hydrogen-burning
stars above
such mass.
 The companion could, however, be a cool WD.
Alternatively, reflection of the WD light by a planet  with
$R_2> 0.04~R_\odot$ (a ``Neptune''), or by a brown dwarf,
 could give the variation amplitude. The insolation temperature
is 1200--2300~K, i.e. this would be another hot planet.
If the albedo is low (giving a temperature close to 2300~K)
and the planet radius
is somewhat larger, thermal re-radiation can also produce the
observed
modulation.
For higher albedos, reflection and thermal re-rediation can
contribute
comparably, or reflection will dominate.
 Interestingly, \citet{ostensen11}, when modeling the
spectrum of this WD, noted its
extremely broad line cores. If broadened by rotation, they imply
rotation with
$v \sin i = 1500$~\kms, about half the break-up speed, which would
mean
 a period of $\sim 40$~s, yet only the 2.24~hr period
appears in the power spectrum. \citet{ostensen11} proposed,
alternatively,
unresolved Zeeman splitting as a possible explanation of the
broad line
cores. In summary, in this WD the modulation can be explained by
rotation
with dichroism
(somewhat unlikely, but would explain also the line profiles),
rotation plus ISM accretion, beaming due to
a WD companion, and reflection or thermal re-radiation from a
brown dwarf or
a hot planet.

\subsection{KIC~11514682 = J19412+4925}
This DA WD has
a mass in the range $M\approx 0.45-0.6~$ M$_\odot$.
 We detect in its light curve a small but significant,
$A=62$ ppm semi-amplitude, modulation with a period of 9.9~d.
It is too hot for spots, but rotation
combined with an uneven surface brightness due
to magnetic dichroism (with $B\sim 35$~kG) or to ISM accretion
and magnetic
channeling (with $\Delta L_{\rm acc}\sim 2\times
10^{29}$~erg~s$^{-1}$)
are viable explanations. For the ISM accretion to work, $n_{\rm
ism}$ would need
to be large and $v_{\rm ran}$ small, given that the WD mass is on
the low side.
Beaming induced by a companion brown dwarf with mass
at the hydrogen-burning limit,
$M_2\approx 0.07~$M$_\odot$, is possible but {\it a priori}
unlikely
because of the rarity of such brown-dwarf companions.
For reflection from a sub-solar companion, the companion
radius would need to be $R_2>0.25 R_\odot$, which is
larger than any known planet.
For an
albedo $D=0.8$, a companion's
insolation temperature would be only $500$~K, making thermal
re-radiation
also a non-option.
Thus, in this case, rotation, or beaming due a sub-stellar
companion, are the possible
alternatives.

\subsection{KIC~11604781 = J19141+4136} In this $\approx 0.6-1.0
$ M$_\odot$ WD we
detect a 4.88~d periodicity with
a semi-amplitude of 0.2 per cent. This is
 the largest amplitude we find, and it
approaches the smallest periodic variation amplitudes
($\gtrsim 0.3$~per cent)
 detected in ground-based observations of the magnetic
WD sample of \citet{brinkworth13}. This is also
the coolest ($T_{\rm eff}=9,100$~K) WD in the sample (it was
selected solely
based on reduced proper motion, rather than UV excess). The same
periodicity that we detect was
found and noted by \citet{ostensen11}. The periodicity
could arise through rotation combined with cool magnetic spots,
dichroism (a rare $B\sim 2$~MG field would be required), or ISM
accretion
(a moderate accretion rate, $\Delta L_{\rm acc}\sim
6\times 10^{28}$~erg~s$^{-1}$, is sufficient due to the
relatively
low luminosity of the WD).
The variation amplitude,
if caused via beaming, would require orbital motion
with $v\sin i= 600$~km~s$^{-1}$, implying an unseen $M>120
$M$_\odot$
black hole companion, which is unrealistic.
 Reflection by a companion would require a companion radius
$R_2>1.3~R_\odot$,
i.e. a sun-like star that would overwhelm the WD spectrum in the
optical.
In this case, WD rotation appears to be the only option to
explain the
photometric variations, through cool convection-inhibiting spots,
hot ISM-accretion spots, (less likely) magnetic dichroism in a
strong field, or (as viable also for all of the variable WDs) UV line 
absorption plus optical fluorescence. 

\section{Discussion}

 We have analyzed
a small sample of 14 WDs
observed by {\it Kepler} and studied by \citet{ostensen10,
ostensen11}.
The sample selection methods
suggest that most, if not all, are normal WDs, rather than WDs
with
strong surface magnetic fields. We have detected 2~hr to 10-day
timescale
photometric variability
in seven of the WDs, at amplitude levels lower than those that
could be seen
previously in WDs. In five of the seven cases the periodicity
is unambiguous, in one case it is highly significant, and in one it is tentative.
The variability could arise from WD rotation combined with
non-uniform
surface emission,
be it due to cool magnetic spots, hot spots from ISM material
channeled to magnetic poles, magnetic dichroism, or perhaps non-uniform UV 
line opacity and optical fluorescence. In five
cases, the WDs
are too hot for convection-inhibited cool
magnetic spots, and in five cases the high
magnetic fields required by magnetic
dichroism are rather unlikely to be present. The ISM accretion
rates
required to reproduce the observed amplitudes are quite high, but
not
impossible. The fluorescence option looks promising, especially in view
of the similar incidence,  of UV line absorption in WDs on the one hand,
 and of variability in 
our sample, on the other hand. 
Accurate measurements of the WD masses, magnetic fields,
random velocities,
ISM densities
along the line of sight, and UV spectra, could permit a clearer test of the
rotation
option.

In five of the cases, we
cannot rule out the possibility
that the variability results from beaming due to orbital motion
caused by cool, low-mass, companion WDs, or by reflection from
cool orbiting
giant planets or brown dwarfs. In two of those five cases thermal
re-radiation
by the substellar companion is also possible. The
WD-companion-beaming
option  can be tested via followup observations in search of
the large radial-velocity variations expected in the primary WD
spectrum.
The WD companion possibility
is unlikely, certainly for more than one or two of the WDs,
 given the observed statistical rarity of close WD
pairs.
Similarly, it is unlikely that more than one or two of 14 WDs
would
have brown-dwarf companions -- such a fraction is far above the
$\lesssim 2\%$
fraction from
direct near-IR searches for brown dwarfs \citep{girven11}.
However,
planets of sub-Jupiter to Jupiter size, if they are abundant at
these
orbits around WDs, could explain three or four cases.


\begin{figure}
  \centering
\includegraphics[width=\linewidth]{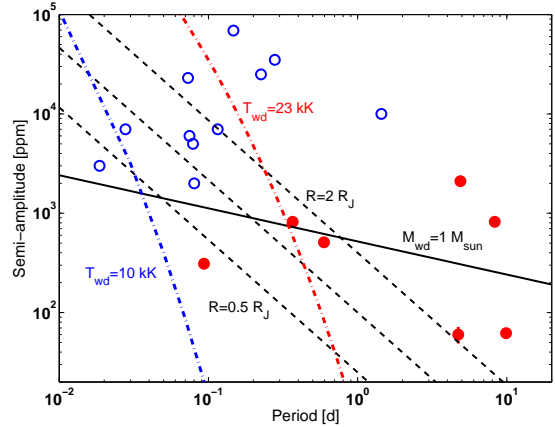}
  \caption{Variation amplitude vs. period for white dwarfs with measured
periods.
Red filled circles are from this study, blue empty circles are from the sample
of
magnetic WDs studied by
\citet{brinkworth04, brinkworth05, brinkworth13} plus the WD from 
\citet{holberg11}. Curves are the
maximum possible amplitudes in several models. Solid black line: 
beaming due to reflex motion 
induced by a 1~M$_\odot$ (e.g. a WD) companion. Dashed black lines: 
reflection by a substellar companion with
0.5, 1.0, and 2 Jupiter radius. Dash-dot curves: thermal re-radiation
from a 2-Jupiter-radius substellar companion, for two WD temperatures,
one (red) characteristic of our sample, and one (blue) 
of the Brinkworth et al. sample's temperatures. 
All models assume an $M_{\rm wd}=0.6~$M$_\odot$ WD and 
 $\sin i =1$. 
}
  \label{fig:peramp}
\end{figure}

Figure~\ref{fig:peramp} shows the seven periods and amplitudes we
have measured,
along with the ground-based measurements by 
\citet{brinkworth04, brinkworth05, brinkworth13}, for 10 
magnetic WDs with fairly 
precise rotation periods (out of 23 WDs monitored by them,
14 of them with sensitivity to periods of up to a week).
Also plotted is a point for the $T=34,000$~K DA WD BOKS53856, for which 
\citet{holberg11} have 
measured, using {\it Kepler}, amplitude $A=2.5\%$  variations 
with a period $P=0.255$~day.
In the rotation models, one
does not
obviously expect a relation between period (which reflects
rotation rate)
and photometric amplitude
(which indicates the degree of surface heterogeneity), unless the magnetic 
field strength correlates with period.
In the beaming scenario and in the planet/brown-dwarf
reflection/re-radiation
 scenario, longer periods, corresponding to larger separations,
 would anti-correlate with amplitude. We include in Figure~\ref{fig:peramp} 
several curves showing the maximal variation amplitudes as a function
of period for these scenarios, for some combinations of parameters.
A 0.6~M$_\odot$ WD and $\sin i =1$ is assumed in every curve.
 The
line for the beaming model assumes a $M_2=1$~M$_\odot$ 
companion (e.g. an unseen cool WD of this mass). The lines for reflection
by a substellar companion are for 
the maximal albedo, $D=1$, and for three possible companion radii, as marked.
The thermal re-radiation curves are calculated for the {\it Kepler} bandpass,
assuming the maximal $R=0.2 R_\odot$ companion radius,
for two WD effective temperatures, $T_{\rm wd}=10$~kK 
charactersitic of the \citet{brinkworth13} sample, and   
$T_{\rm wd}=23$~kK 
typical of our sample. 

The figure illustrates, for each WD,
 what are the allowed mechanisms
from among these options (as discussed in more detail for each case in
Section~\ref{sec:indiv}).
Interestingly, the
Brinkworth et al.
WDs, which have larger amplitudes than our sample (only such
large
amplitudes
were detectable by their ground-based data),
have shorter periods, typically in the
sub-day range. Rather than spottedness (the favoured explanation
by
Brinkworth et al.), we see that reflection or re-radiation
from planets could be the cause
for some of those WDs below the maximum-amplitude vs. period relations, plotted 
in Figure~\ref{fig:peramp}.  It is also remarkable that Brinkworth et al. found
clear periodic variations in 10 out of 23 WDs they studied, again a 
$\sim 50\%$ fraction. If fluoresence is the cause in that sample as well,
the larger variation amplitudes could result from the stronger magnetic fields
of their sample's WDs, which could lead to stronger channeling of the 
accretion flow, and hence to a more non-uniform UV opacity and 
optical re-emission. 

Some cases appear to defy most of the proposed explanations.
\citet{brinkworth13} have detected, in the WD PG1658+441,
rather large-amplitude (2\%)
variability, with a probable period
between 6~hr and 4~d (not included in Figure~\ref{fig:peramp} because
of its uncertain period).
This is a DA-type WD with $T_{\rm eff}=30,500$~K,
i.e. too hot for spots, and with
$B=3.5$~MG, a field an order of magnitude too low
for magnetic dichroism-induced
variability of this amplitude.
Furthermore,
infrared observations have excluded also the presence of
companions
above 10 Jupiter masses \citep{hansen06, farihi08}.
Very similarly, the WD BOKS53856 ($T=34,000$~K, $P=0.255$~day, $A=2.5\%$,
plotted) likely has only $B\sim 350$~kG.
\citet{holberg11} have proposed, as an explanation,
 rotation combined with chemical 
abundance-induced temperature variations over the stellar surface,
likely associated with ongoing accretion (not dissimilar to our proposed
mechanism of fluorescence).  

If some of the seven cases
 of periodic variations in our sample result from WD rotation
(with the modulation due to spots, dichroism, or accretion),
then our results are among
the few accurate measurements of rotation periods in common (DA
and DB)
WDs that are
non-pulsating and not necessarily magnetic.
This is made possible by the photometric precision of
{\it Kepler}, which permits detecting variations
 at much subtler levels than previously possible.
While we have measured periods for only five to seven WDs, the periods
are all relatively long,
 and
of the same order as the rotation periods
previously found for more peculiar WDs. This may hint that slow WD
rotation is
 the norm, and fast-rotating WDs may indeed be rare among the
full WD
population.

Alternatively, if the correct interpretation of our results
is that of reflection or re-radiation
 from planets, then hot giant planets are extremely
common around WDs, even if this interpretation is true for only a
few of the
seven dwarfs.
 Indeed, considering that the detection limits for
several of the WDs without detected variability are above some of
the detected
amplitudes, a majority of WDs could have such planets. Such a
discovery,
while dramatic, would not be completely unexpected, given the
recent  accumulating
evidence for the presence, around many WDs, of dust-and-gas 
disks, and ongoing accretion of rocky debris 
(see Section~\ref{sec:intro_companions}).
If snowline-region Neptunes
exist around many stars, as indicated by microlensing planet
surveys \citep{gould10, cassan12, shvartzvald14}
and inward migration of such planets is frequent in the
protoplanetary phase
(e.g. \citealt{trilling98, ida08}), it is conceivable that
post-stellar migration of
surviving
outer planets could frequently occur around WDs.
In this process, the observed disks around WDs
might play a similar role to that of proto-planetary
disks around young stars.

To summarise, we have reported the first detections of periodic 
modulations in WDs at amplitudes $\lesssim 10^{-3}$. More work is required
to understand this uncharted territory.
Our results highlight the need and the
possibility, already being pursued,
 of obtaining such measurements for larger samples of WDs
as part of the K2 {\it Kepler}
Mission extension. Even larger WD samples with time-domain information 
will come from the {\it Gaia}, TESS, and PLATO missions.
With such samples, it will be possible to
search for relations between periodic variations
and other instrinsic properties of WDs. This should help
discriminate among the diverse explanations for the intrigung
periodic variability, seen in the small sample studied here.

\section*{Acknowledgments}
We thank J. Farihi, B. Gaensicke, T. Marsh, 
E. Nakar, and the anonymous 
referee for useful advice and input. E. Goldstein is thanked
for his significant contributions to the later versions of the paper.
D.M and T.M  acknowledge support from the Planning and Budgeting
Committee's
Israeli Centers of Research Excellence (I-CORE, grant
No.\,1829/12). T.M. acknowledges support from
the European Research Council under the EU's Seventh Framework
Programme (FP7/(2007-2013)/ERC Grant Agreement No.\,291352),
and by the Israel Science Foundation (grant No.\,1423/11).
This research has made use of the NASA Exoplanet Archive, which
is
operated by the California Institute of Technology, under
contract with the National Aeronautics and Space Administration
under the Exoplanet Exploration Program. All of the data
presented in this paper were obtained from the Mikulski Archive
for Space Telescopes (MAST). STScI is
operated by the Association of Universities for Research in
Astronomy, Inc., under NASA contract NAS5-26555. Support for MAST
for
non-HST data is provided by the NASA Office of Space Science via
grant
NNX09AF08G and by other grants and contracts.

\bibliography{wdbib} \bibliographystyle{apj}

\end{document}